# Data Stealing Attack on Medical Images: Is it Safe to Export Networks from Data Lakes?

Huiyu Li[1], Nicholas Ayache[1], Hervé Delingette [1]

[1] Inria, Université Côte d'Azur, France
huiyu.li@inria.fr

**Abstract.** In privacy-preserving machine learning, it is common that the owner of the learned model does not have any physical access to the data. Instead, only a secured remote access to a data lake is granted to the model owner without any ability to retrieve data from the data lake. Yet, the model owner may want to export the trained model periodically from the remote repository and a question arises whether this may cause is a risk of data leakage. In this paper, we introduce the concept of data stealing attack during the export of neural networks. It consists in hiding some information in the exported network that allows the reconstruction outside the data lake of images initially stored in that data lake. More precisely, we show that it is possible to train a network that can perform lossy image compression and at the same time solve some utility tasks such as image segmentation. The attack then proceeds by exporting the compression decoder network together with some image codes that leads to the image reconstruction outside the data lake. We explore the feasibility of such attacks on databases of CT and MR images, showing that it is possible to obtain perceptually meaningful reconstructions of the target dataset, and that the stolen dataset can be used in turns to solve a broad range of tasks. Comprehensive experiments and analyses show that data stealing attacks should be considered as a threat for sensitive imaging data sources.

## 1 Introduction

Machine learning-based methods have the potential to revolutionize the field of medical data processing. To further develop this promising technology, a growing number of medical data warehouses or data lakes have been built within major hospitals or health organisations. With those infrastructures, the access of health data such as medical images or health records is heavily restricted and regulated, and only a remote access to the training and test data is granted to data scientists sitting outside those organizations. Any leakage of privacy sensitive medical data from those data lakes represents a serious threat to the reputation of the health organization holding the data lake, and it may also be used by cybercriminals to earn money through ransoms, or to cause harms [12].

**Table 1.** Data attacks extracting information about the training data. (Top) Attacks extracting data properties; (Bottom) Attack recovering the training data.

| Attack | Adversary Knowledge | | | | |
|---|---|---|---|---|---|
| | Training Data | Model | | Output | |
| | | *Architecture* | *Parameters* | *Final* | *Intermediate* |
| Property inference attack [10] | ✗ | ✗ | ✓ | ✗ | ✗ |
| Reconstruction attack [10] | ✗ | ✓ | ✓ | ✓ | ✗ |
| Membership inference attack [10] | ✗ | ✓ | ✓ | ✓ | ✗ |
| Inverting visual representations [6] | ✗ | ✓ | ✓ | ✗ | ✓ |
| Model inversion attack [10] | ✗ | ✓ | ✓ | ✓ | ✗ |
| Inverting gradients [8] | ✗ | ✓ | ✓ | ✗ | ✗ |
| **Data stealing attack** | ✗ | ✓ | ✓ | ✗ | ✓ |

A number of AI-related cyber-attacks such as adversarial attacks [12] have been studied in the literature. In this paper, we are interested in attacks targeting the extraction of information from images in the training set. Indeed, previous studies on possible cyber-attacks have shown that trained models encapsulate some information about the training data, thus making them vulnerable to privacy attacks. In Table 1, such attacks are listed including the specific knowledge of the attacker about the trained model or the output of the model on the training data. The first group of attacks such as property inference [7] or reconstruction attacks [15] tries to retrieve some partial information about the training data. Membership inference attacks [9] identify whether a data sample is present in the dataset.

The second group of attacks is aiming to reconstruct partially or entirely images in the training set from the knowledge of the complete model and some model output. Early work aimed at inverting visual representations [6] from some intermediate output of a neural network. This model inversion is done by training a neural network on a known image dataset similar to the one to be recovered. However, this method leads to image reconstructions of limited quality and more sophisticated model inversion attacks have been recently proposed [4] based on GANs. In this case, the discriminator is trained to distinguish between fake and real images from a known prior image set. This approach creates realistic but not exact copies of the original images. Finally, our study is also related to inverting gradient methods [8] that try to recover input images from model parameter gradient that leak during training in federated learning framework.

In this paper, we introduce a new attack, the data stealing attack, allowing an attacker to appropriate training data from a remote data lake or in a federated learning setting. This attack is solely based on the export of a trained model and makes both limited and realistic assumptions. It consists in training an algorithm to perform lossy image compression and then to hide the image compression codes and the decoder into the exported neural network. Thus, the attacker can regenerate the training set images with high perceptual quality outside the data lake by applying the decoder on the image codes. Besides, we show that a dedicated branch of the compression network can still solve a utility task such as segmenting an image, thus making it diffi-cult to detect the nature of attack. To the best of our knowledge, this is the first work using image compression learning to develop such type of data attacks. Furthermore, we show that such attacks may be realistically deployed in the sensitive context of medical imaging.

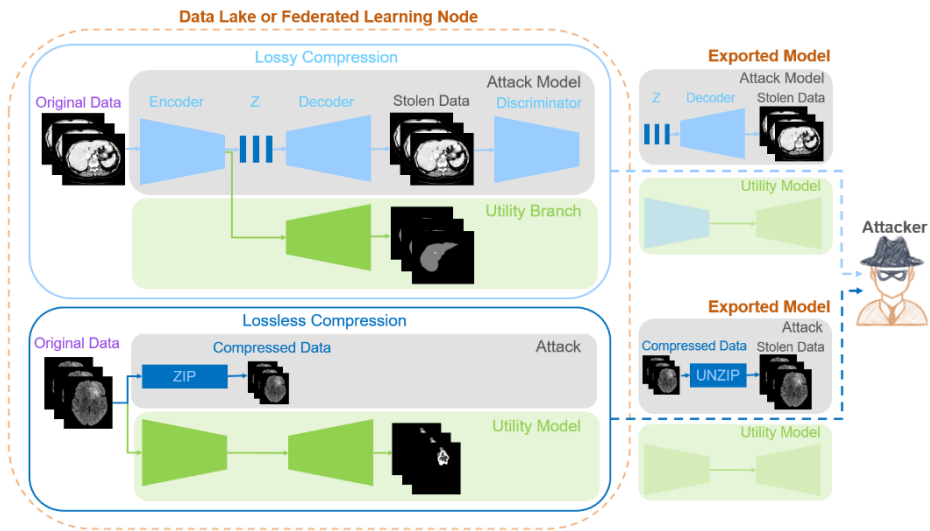

**Fig. 1.** Overview of the proposed data stealing attack method.

## 2 Data Stealing Attack

### 2.1 Attack Strategy

**Attack Assumptions.** We consider a practical setting where a remote user is given access to sensitive image data that are stored in a secured data lake as shown in Fig. 1. This configuration is commonly encountered for instance when data scientists are remotely accessing health datasets in hospitals without the ability to retrieve locally the data for regulatory reasons. This is also the situation encountered in *federated learning* where the data lake corresponds to a participating node in a centralized or decentralized architecture. Inside the data lake the remote user has free access to some original imaging data that can serve to solve a *utility task*. This task may be for example solving an image segmentation problem. Eventually, the remote user asks the data owner to retrieve the trained model in order to exploit it for its own purposes. In the case of federated learning, the locally trained model is periodically sent to a central server or another participating node in order to be aggregated into a global model. Without sharing any data outside the data lake, this seems to be a robust privacy preserving framework, but what if an honest-but-curious remote user acts as an attacker? The possible motivations of an attacker to create a data breach may be to reuse the data for solving other tasks, to cause harm to the reputation of the data owner, or to ransom the owner. The attack may also take place after the attacker has stolen the identity of a trusted honest user.

**Attack Principle.** The attack consists in exporting a neural network from the data lake that solves the utility task but that also contains image codes allowing the attacker to reconstruct with or without losses images stored in the data lake. A limiting factor of the attack is the size of the exported network, since exporting a very large network may be suspicious to the data owner. Therefore, the objective of the attacker is to maximize the number of stolen images while minimizing the disk size of the exported model.

To tackle this trade-off, the attacker may adopt either lossy or lossless compression approaches. In lossless compression (see Fig. 1 bottom), the attacker can apply a standard compression tool, such as *ZIP*, *or RAR* on images in the data lake and then store the compressed images inside the exported utility model. Yet, lossless image compression usually produces restricted compression ratio thus potentially limiting the number of images that can be stolen.

An interesting alternative is to develop lossy compression algorithms (see Fig. 1 top) reaching low bits per pixel but requiring a domain-specific encoder and decoder. For that purpose, we adopt in this paper the generative image compression model developed in [14] that combines GAN with learned compression techniques. It includes an encoder that transforms an image x into its latent code $y = E(x)$ and a decoder or generator which transforms the code y into an approximation of the original image, $x' = G(y) \approx x$ and a discriminator $D(x')$ to decide if the generated image is real or fake. In addition, a utility model solving for instance an image segmentation task must be devised in order to convince the data owner that the exported model is effective. To create a light utility model taking a limited amount of disk space, the attacker can use the encoder of the generative compression model as the feature extraction network, and train a decoding branch that is specialized in the utility task. In this case, the exported model includes the generator G (a.k.a the decoder) and the image codes generated by the encoder that can be hidden in the neural network. The attacker can then generate the images outside the data lake by applying the decoder on the image codes.

**Case of Centralized Federated Learning.** In this setup, several aggregation steps are iteratively applied to send a local model from each participating node to a central server. Therefore, if the attacker controls the central server, each aggregation step may be an occasion to steal some data in each local model. Besides, the server may send to each node an image encoder $E(x)$ while each node may send to the server only a set of image compression codes hidden as network weights. The server can then use a decoder to gather a large number of lossy reconstructed images from each node.

**2.2  Attack Implementation Training Pipeline.**

In the lossy compression case, the attacker starts to train the generative image compression model composed on an encoder, generator, and discriminator networks. The input of the encoder is a 256 × 256 image with three channels suitable for compressing color images, but to handle volumetric medical images, specific preprocessing steps are detailed in section 3.1. Once the compression model is trained, all images are encoded. Then the attacker freezes the parameters of the encoder network and begins to train the utility branch to solve the utility task. It is sufficient for the attacker to obtain reasonable results for that task to convince the data owner to export the trained network.

**Hiding Image Codes in Network Weight Files.** In both cases, the data stealing attack assumes that image codes are hidden in network weight binary files. Indeed, those weights are commonly saved in HDF5 file formats where the weights of each layer are stored in a dictionary. Image codes may then be added as entries to the dictionary with dedicated keys making them easy to retrieve.

## 3  Experiments

### 3.1  Datasets and Models

We evaluate the effectiveness of our attack model on two public datasets of medical images. The former is the MICCAI 2017 Liver Tumor Segmentation (LiTS) Challenge dataset [2] that contains 130 CT cases for training and 70 CTs for testing. In this dataset, the utility task is to segment the liver parenchyma in a supervised manner. The second dataset is the BraTS 2021 challenge dataset [1] which includes 1251 skull-stripped brain images with multiple MR sequences for training and 219 cases for validation. The utility task is to segment the whole tumor based on FLAIR MR sequences.

On the LiTS (resp.BraTS 2021) dataset, we randomly partition the training set into 104/13/13 (resp. 1000/126/125) images that are used for training, validation, and testing of the utility task. Also, for testing the lossy compression network, we use the 70 (resp. 219) test images in the LiTS (resp. BraTS) dataset.

**Pre and Post-processing.** Each slice of the LiTS CT images is of size 512×512 whereas the input size of the encoder network is 256 × 256 × 3. Two different approaches were tested corresponding to two different cost-quality compromises. The first method (Low) is to downsample each slice by a factor of 2 while the second (High) is to decompose each 512×512 slice into 3×3 overlapping patches that are separately encoded. Thus, the latter requires 9 times more image codes than the former to reconstruct an image. In the BraTS dataset, edge padding is applied since the slice resolution is only 240 × 240 pixels. Finally, a minmax intensity normalization is applied on the whole image, and each slice is surrounded by its upper and lower slices to fill the three input channels. For post-processing, the image intensity is mapped back to its original minimum and maximum range and upsampling with bilinear image blending is used to reconstruct the original slices for Low/High slice sampling.

**Table 2.** Fidelity & compression results on the LiTS and BraTS datasets. 'BPP$_{input}$/ BPP$_{comp}$': bit per pixel of input / compressed data, 'P$_{ratio}$': practical ratio.

| Input | BPP$_{input}$ | BPP$_{comp}$ ↓ | PSNR↑ | MS_SSIM↑ | P$_{ratio}$ ↓ |
|---|---|---|---|---|---|
| High$_{Training}^{LiTS}$ | 17.858±1.791 | 0.221±0.053 | **40.322±0.793** | 0.992±0.002 | 0.168±0.019 |
| Low$_{Training}^{LiTS}$ | 17.858±1.791 | **0.097±0.027** | 38.193±0.444 | 0.987±0.002 | **0.017±0.002** |
| High$_{Testing}^{LiTS}$ | 16.289±1.899 | 0.125±0.024 | 40.306±1.096 | **0.995±0.001** | 0.185±0.023 |
| Low$_{Testing}^{LiTS}$ | 16.289±1.899 | 0.145±0.029 | 33.424±1.021 | 0.981±0.004 | 0.021±0.002 |
| High$_{Training}^{BraTS}$ | 3.801±0.285 | **0.241±0.100** | 37.842±1.687 | 0.996±0.001 | 0.395±0.024 |
| High$_{Testing}^{BraTS}$ | 3.926±0.282 | 0.250±0.100 | 36.070±2.280 | 0.995±0.001 | **0.387±0.025** |

**Image Compression and Utility Models.** Following [14], to speed-up training, the image compression model is first trained with rate and distortion losses only, then with all losses in a second stage. With lossy compression networks, the utility task is solved with a Utility Branch (UB) model connected to the last layer of the image encoder network. In that case, the dedicated branch corresponds to a simplified version of the 2D CNN segmentation network in [13]. When lossless compression is chosen, we train from scratch an off-the-shelf model [3] coined as Public Utility (PU) model in the remainder. All models are optimized with Adam [11] and training continues until the validation loss has converged.

### 3.2 Effectiveness of Data Stealing Attacks

**Compression-fidelity Compromise.** We evaluate the ability of stealing medical images with the data stealing attack strategy based on two datasets. First, Table 2 reports the trade-off between image fidelity and compression ratio. The practical ratio is the ratio of the disk space needed to store the image codes of a volumetric image (lossy compression) to the disk space to store the ZIP compressed image (lossless compression). On the LiTS dataset, the low slice sampling approach leads to image codes 60 times smaller than an image compressed by ZIP. The high slice sampling approach requires 10 times more disk space but leads to higher image fidelity. On the BraTS dataset, however, the lossy compression gain is far smaller probably due to the large amount of uniform background in the original images. Good fidelity reconstruction is obtained with a PSNR of nearly 40 although values higher than 60 are considered as high quality for 12-bit images [5]. A visual comparison between original and reconstructed images is available in Fig. 2 for both training and test sets. Note that in this specific case, the trained generative compression model is typically applied on training data not testing data.

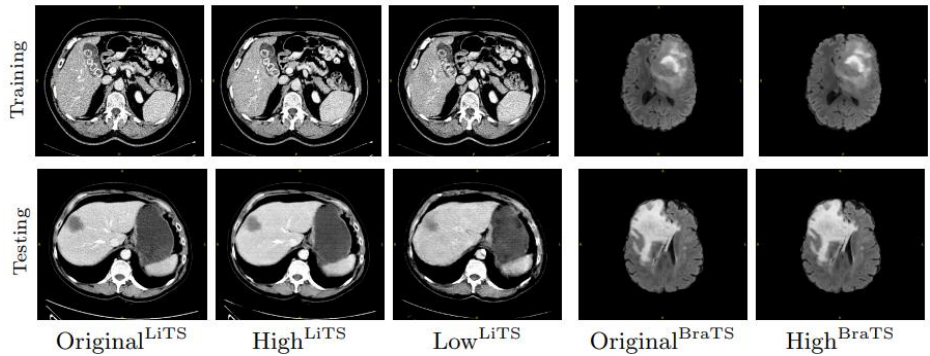

**Fig. 2.** Lossy image reconstructions on training and testing images from the LiTS and BraTS datasets. From left to right: original data, highly sampled reconstructions, under sampled reconstructions (LiTS dataset only).

**Utility Task Performances.** In Table 3, we report performances of the two utility models, branch (UB) and public (PU) models to solve the liver (resp. whole tumor) segmentation on the LiTS (resp. BraTS) dataset. Those models are trained on both the original images in the data lake and the lossy reconstructed (or stolen) version of the training set. Seven commonly used metrics are listed including Dice score, volumetric overlap error (VOE), relative volume difference (RVD), average symmetric surface distance (ASSD), maximum surface distance (MSD), and root means square symmetric surface distance (RMSD) [13]. The same unseen test image set is used for the three utility models. We see that the performances of the public model are the same on the original and stolen data, showing that the slight degradation of image quality due to image compression does not impact its generalization ability. The branch model is clearly less efficient since it is based on a frozen encoder branch. Yet, it leads to an average 0.88 Dice score, which makes it a plausible network to solve this task.

**Table 3.** Utility task results on LiTS testing dataset. 'UB': the utility branch model, 'PU': the public utility model, 'stolen': the stolen dataset.

| Methods | Dice↑ | VOE↓ | RVD↓ | ASSD↓ | MSD↓ | RMSD↓ |
|---|---|---|---|---|---|---|
| UB | 0.866±0.047 | 0.233±0.070 | 0.020±0.118 | 4.078±1.256 | 40.104±14.654 | 6.274±1.896 |
| PU | 0.928±0.064 | 0.128±0.100 | -0.001±0.049 | 2.544±3.051 | 45.274±35.978 | 5.476±6.327 |
| PU$_{stolen}$ | 0.926±0.073 | 0.131±0.112 | -0.015±0.060 | 2.832±3.987 | 47.236±36.360 | 5.933±7.462 |

**Trade-off between Network Size and the Number of Stolen Images.** In Table 4, we estimate the disk size of three exported models (checkpoint files) involved in a data stealing attack on both the BraTS and LiTS datasets trying to steal 100 original images. In lossy compression, the decoder is very large (600 MB) but the generated image code per image is small: in average 2.2MB (resp. 22MB) for low (resp. high) slice sampling for LiTS CT dataset, and 0.9MB for the BraTS dataset. With lossless compression, there is no need to export the decoder but the ZIP compressed images are fairly large to store: in average 134MB for each CT scan in LiTS and 2.3MB for BraTS. The branch utility model has negligible disk size and the results in Table 4 suggest that an attacker willing to optimize the exported model disk size, would pick a lossy compression for CT images and lossless compression for MR skull-stripped images.

### 3.3 Mitigation of Data Stealing Attacks

We have shown that in the data stealing attack, the attacker exports a network that solves a utility task with reasonable performances. To detect a potential data breach, the data owner may want to check the size of the exported model considering large models as suspicious. In our test, the compression decoder is fairly large (598MB) but has typically a similar size as a backbone such as VGG16 (576MB). It is probably possible to largely decrease the disk size of such decoder by using for instance network quantization, or drop-out. Furthermore, in the case of centralized federated learning, the decoder need not be exported and therefore the size of the exported model may be less than 100MB to steal around 50 CT images. Therefore, a more robust mitigation to this type of attack is probably to certify that the code running in a data lake guarantees data privacy.

**Table 4.** Disk size needed to steal 100 images with various attack strategies. 'D': the decoder of attack model, 'UB'/'PU': the utility branch and public utility models, 'High/Low/ZIP': lossy or lossless compressed codes.

| Dataset | Disk Size (MB) | | | | |
|---|---|---|---|---|---|
| | D | D + UB | D + UB+ $100 \cdot \text{High}_{\text{Training}}$ | D + UB+ $100 \cdot \text{Low}_{\text{Training}}$ | PU+ $100 \cdot \text{ZIP}_{\text{Training}}$ |
| LiTS | 598 | 601 | 2800 | **828** | 13466 |
| BraTS | 598 | 601 | 692 | / | **260** |

## 4 Conclusion

In this paper, we have introduced a novel attack aiming to steal training data from a data lake or from participating nodes in federated learning. An attacker proceeds by using a learned generative lossy image compression network and exporting a decoder together with image codes. An alternative for stealing image annotation masks for instance is to use lossless compression with standard tools. We have shown that such attacks are feasible on two medical imaging datasets with a trade-off between the size of the exported network and the number of stolen images.